\documentclass[conference]{IEEEtran}
\IEEEoverridecommandlockouts
\usepackage{cite}
\usepackage{amsmath,amssymb,amsfonts}
\usepackage{algorithmic}
\usepackage[detect-none]{siunitx}
\usepackage{graphicx}
\usepackage{textcomp}
\usepackage{xcolor}
\usepackage[a4paper, total={184mm,239mm}]{geometry}
\def\BibTeX{{\rm B\kern-.05em{\sc i\kern-.025em b}\kern-.08em
    T\kern-.1667em\lower.7ex\hbox{E}\kern-.125emX}}
    
\usepackage{booktabs}
\usepackage{graphics}  
\usepackage{amsmath}            
\usepackage{amssymb} 
\usepackage{url}

\def\figref#1{Fig.~\ref{#1}}
\def\tabref#1{Tab.~\ref{#1}}
\def\eqref#1{Eq.~(\ref{#1})}

\begin{document}

\title{Fully On-board Low-Power Localization with Multizone Time-of-Flight Sensors on Nano-UAVs\\

\vspace{-0.5cm}}
\author{
\IEEEauthorblockN{Hanna M\"uller\IEEEauthorrefmark{1}\IEEEauthorrefmark{3}, Nicky Zimmerman\IEEEauthorrefmark{2}\IEEEauthorrefmark{3}, Tommaso Polonelli\IEEEauthorrefmark{1}}
\IEEEauthorblockN{Michele Magno\IEEEauthorrefmark{1}, Jens Behley\IEEEauthorrefmark{2}, Cyrill Stachniss\IEEEauthorrefmark{2}, Luca Benini\IEEEauthorrefmark{1}}

\IEEEauthorblockA{\IEEEauthorrefmark{1}Integrated Systems Laboratory / Center for Project-Based Learning  - ETH Z\"urich, Switzerland}

\IEEEauthorblockA{\IEEEauthorrefmark{2}Institute of Geodesy and Geoinformation - University of Bonn}

\IEEEauthorblockA{\IEEEauthorrefmark{3} The authors contributed equally to the paper.}

Email: {hanmuell, tpolonelli, mmagno, lbenini}@ethz.ch, nicky.zimmerman, jens.behley, cyrill.stachniss@igg.uni-bonn.de
\vspace{-0.65cm}
}

\maketitle

\begin{abstract}
Nano-size unmanned aerial vehicles (UAVs) hold enormous potential to perform autonomous operations in complex environments, such as inspection, monitoring or data collection. Moreover, their small size allows safe operation close to humans and agile flight. An important part of autonomous flight is localization, which is a computationally intensive task especially on a nano-UAV that usually has strong constraints in sensing, processing and memory. This work presents a real-time localization approach with low element-count multizone range sensors for resource-constrained nano-UAVs. The proposed approach is based on a novel miniature 64-zone time-of-flight sensor from ST Microelectronics and a RISC-V-based parallel ultra low-power processor, to enable accurate and low latency Monte Carlo Localization on-board. Experimental evaluation using a nano-UAV open platform demonstrated that the proposed solution is capable of localizing on a 31.2m$\boldsymbol{^2}$ map with 0.15m accuracy and an above 95\% success rate. The achieved accuracy is sufficient for localization in common indoor environments. 
We analyze tradeoffs in using full and half-precision floating point numbers as well as a quantized map and evaluate the accuracy and memory footprint across the design space.
Experimental evaluation shows that parallelizing the execution for 8 RISC-V cores brings a 7x speedup and allows us to execute the algorithm on-board in real-time with a latency of 0.2-30ms (depending on the number of particles), while only increasing the overall drone power consumption by 3-7\%. 
Finally, we provide an open-source implementation of our approach.
\end{abstract}


\vspace{-0.45cm}
\section{Introduction}
Nano-size unmanned aerial vehicles (UAVs) fit in the palm of a hand, weight only a few tens of grams, and therefore are agile, able to pass through narrow gaps and safe to operate in proximity of humans~\cite{tsykunov2018swarmtouch}. Achieving autonomous flight is one of the most promising and difficult challenges for nano-UAVs, as they need to execute key tasks for autonomous robot navigation such as obstacle avoidance, localization, mapping and path planning~\cite{loquercio2019deep}. However, the restrictions in payload and power consumption pose severe challenges in reaching the autonomy of standard-size drones, as sensing and processing are strongly limited \cite{mcguire2019minimal}.

Focusing on the  localization task, RTK-GPS is commonly used for outdoor scenarios~\cite{schneider16icra}. Instead, in GPS-denied environments, such as in indoor scenarios, onboard localization is challenging, especially on nano-UAVs, as they can only afford to spend around 10-15\% for sensing and processing~\cite{wood2012progress}. Enabling indoor localization on such constrained platforms is pushing researchers to design approaches that are lightweight and efficient --- the most common ones are radio-based localization methods (mostly ultra-wideband (UWB)~\cite{van2020board, niculescu2022iotj}) or approaches that require off-board processing~\cite{simsek2021blur}. However, these approaches have major drawbacks, such as depending on external infrastructure and reliable communication to other nodes or a powerful basestation with a reliable wireless connection for off-board processing~\cite{simsek2021blur}.

A successful and popular localization methodology is map-based, which does not rely on external infrastructure~\cite{moravec1989sdsr,meyerdelius2021aaai,zimmerman2022iros}. To estimate the pose in a given occupancy grid map~\cite{moravec1989sdsr}, it is necessary to have range measurements, which can be obtained by range-sensors such as LiDARs and depth/stereo cameras~\cite{debeunne2020review}. However, these sensors are large, power-hungry, and in the case of stereo cameras, also computationally expensive to process, thus making them unsuitable for most nano-UAVs \cite{mcguire2019minimal}. A promising alternative is recently-emerged multizone time-of-flight (ToF) sensors, which is also suitable for nano-UAVs and already proved to be robust and reliable for obstacle avoidance \cite{mueller2022robust}. The drawback of this sensor is its low element-count, which is sufficient for obstacle avoidance but proves challenging for localization.
\begin{figure}[t]
  \centering
  \includegraphics[width=0.9\linewidth,trim={0 0 0 5cm},clip]{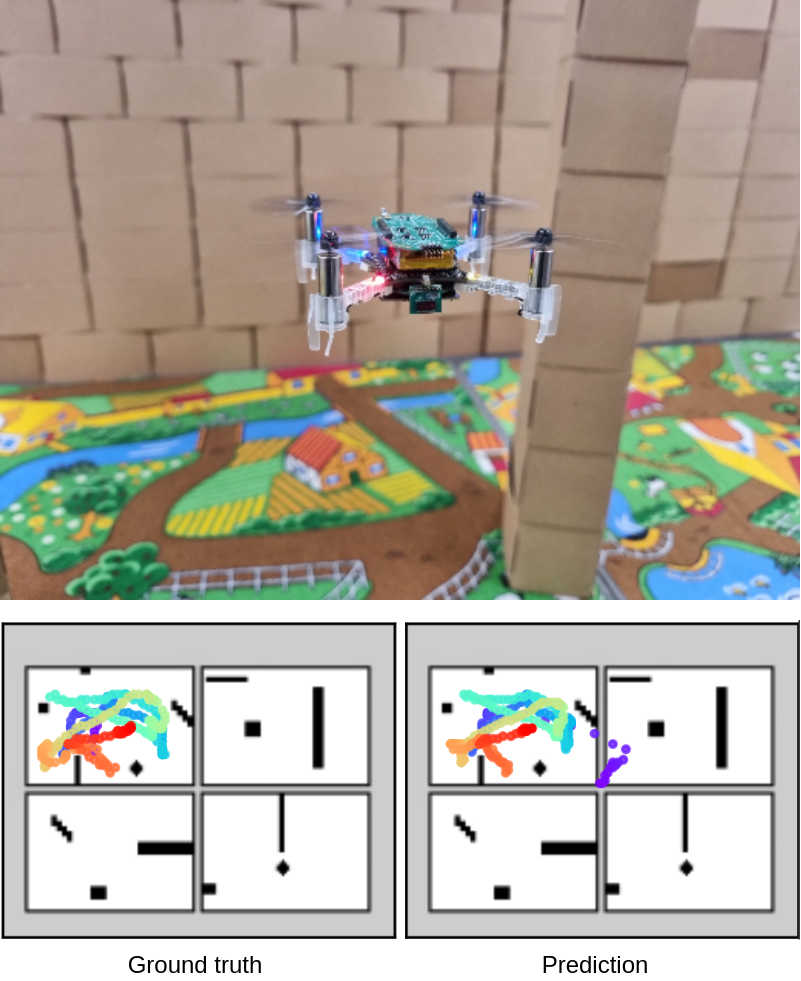}
  \vspace{-0.35cm}
   \caption{Top: the nano-UAV localizing in the maze. Bottom: The ground truth pose and the predicted pose for 4096 particles. The initial prediction starts off in the wrong maze and then converges to the correct pose when more observations become available. The color encodes the time, purple is the start, red the end.}
  \label{fig:motivation}
  	\vspace{-0.75cm}
\end{figure}

This paper proposes a global localization system, based on Monte Carlo localization (MCL) and exploiting novel miniaturized low-power ToF sensors, not requiring previously installed infrastructure. The proposed algorithm is designed to run online on nano-UAV processors. In particular, the paper proposes an algorithm where all the computations are performed on-board on a milliwatt power RISC-V parallel system on chip (SoC), avoiding communication latency and potential issues such as range limits or security risks while significantly improving the latency. To exploit the hardware architecture of the RISC-V-based SoC, this paper presents a parallelized and memory-efficient implementation tailored for the system's computational capabilities. Finally, the proposed approach has been experimentally evaluated in the field, and an open-source implementation will be available. 
The nano-UAV with all sensors and processors mounted is shown in Fig.~\ref{fig:motivation}.
In our experiments, we show that our approach
is able to (i) accurately localize a nano-UAV in a given map, 
using low element-count sensors without infrastructure,  (ii) reduce memory consumption with quantization and lower precision floats without a significant drop in accuracy, (iii) reduce latency by a factor of 7 through parallelized implementation and localize on-board in real-time, and (iv) operate with low power consumption, where sensing and processing only consume below 7\% of the overall power.

\section{Related Work}
Recent literature has demonstrated that sensing and processing on nano-UAVs is strongly limited, therefore, many previous works have proposed solutions for autonomous navigation that only rely on simple state estimation techniques such as an inertial measurement unit and odometry for localization~\cite{mcguire2019minimal, mueller2022robust, palossi_frontnet, niculescu2021improving}. The major drawback of these approaches is their inability to compensate for drift and recover from accumulated errors \cite{mueller2022robust}. 
Most drift-correction approaches use range measurements to anchors with known locations \cite{niculescu2022iotj}. 

In indoor scenarios, the most commonly used technology is UWB, but approaches with Bluetooth or WiFi are also available. They all have disadvantages --- they require line-of-sight between nodes, depend on pre-installed  infrastructure~\cite{niculescu2022iotj} or can only estimate relative position~\cite{van2020board}. The resulting mean localization errors are often over \SI{20}{\centi\meter} (\SI{22}{\centi\meter} in \cite{niculescu2022iotj}, \SI{28}{\centi\meter} in \cite{van2020board}).

In contrast to previous works, this paper focuses on an infrastructure-less approach to globally navigate indoors: a map-based localization approach using particle filters, which was not explored on nano-UAVs until now.

Localization in a given map is an essential capability of most autonomous robot systems, laying the foundation for more complex tasks such as planning and manipulation. Probabilistic approaches provide robust localization and include seminal works such as the extended Kalman filter (EKF)~\cite{leonard1991tra}, Markov localization~\cite{fox1999jair} and particle filters often referred to as MCL~\cite{dellaert1999icra}. These approaches are suitable for localization using range sensors such as 2D LiDARs and sonars, as well as cameras.
Until now, this approach was nearly infeasible on nano-UAVs, due to bulky power hungry sensors, and high computational demands which are hard to satisfy on embedded systems.\\

For both the sensing and the processing challenges, promising hardware recently emerged. Although introduced on the market only recently, lightweight multizone ToF sensors are already working well for obstacles avoidance \cite{mueller2022robust}. 
As for powerful and energy-efficient SoCs for processing, SoCs of the parallel ultra-low power (PULP) family have been employed on drones before. The GAP8 SoC is utilized for corridor~\cite{niculescu2021improving} or person following~\cite{palossi_frontnet}. These approaches use deep learning with quantized models, but they do not venture into float-heavy tasks such as particle-filter localization. 
A novel SoC, GAP9\footnote{\url{https://greenwaves-technologies.com}}, was recently released, which with \SI{0.33}{\milli\watt} per giga operation (GOP) is an order of magnitude more power efficient than GAP8 and most importantly, features increased memory and floating point support.

In this work we combine a miniature multizone ToF sensor with a novel processor to enable on-board infrastructure-less localization in indoor environments with an accuracy that surpasses the state of the art of localization in nano-UAVs with UWB~\cite{van2020board, niculescu2022iotj}.
\vspace{-0.3cm}
\section{System Architecture}
\begin{figure}[t]
  \centering
  \includegraphics[width=1.0\linewidth]{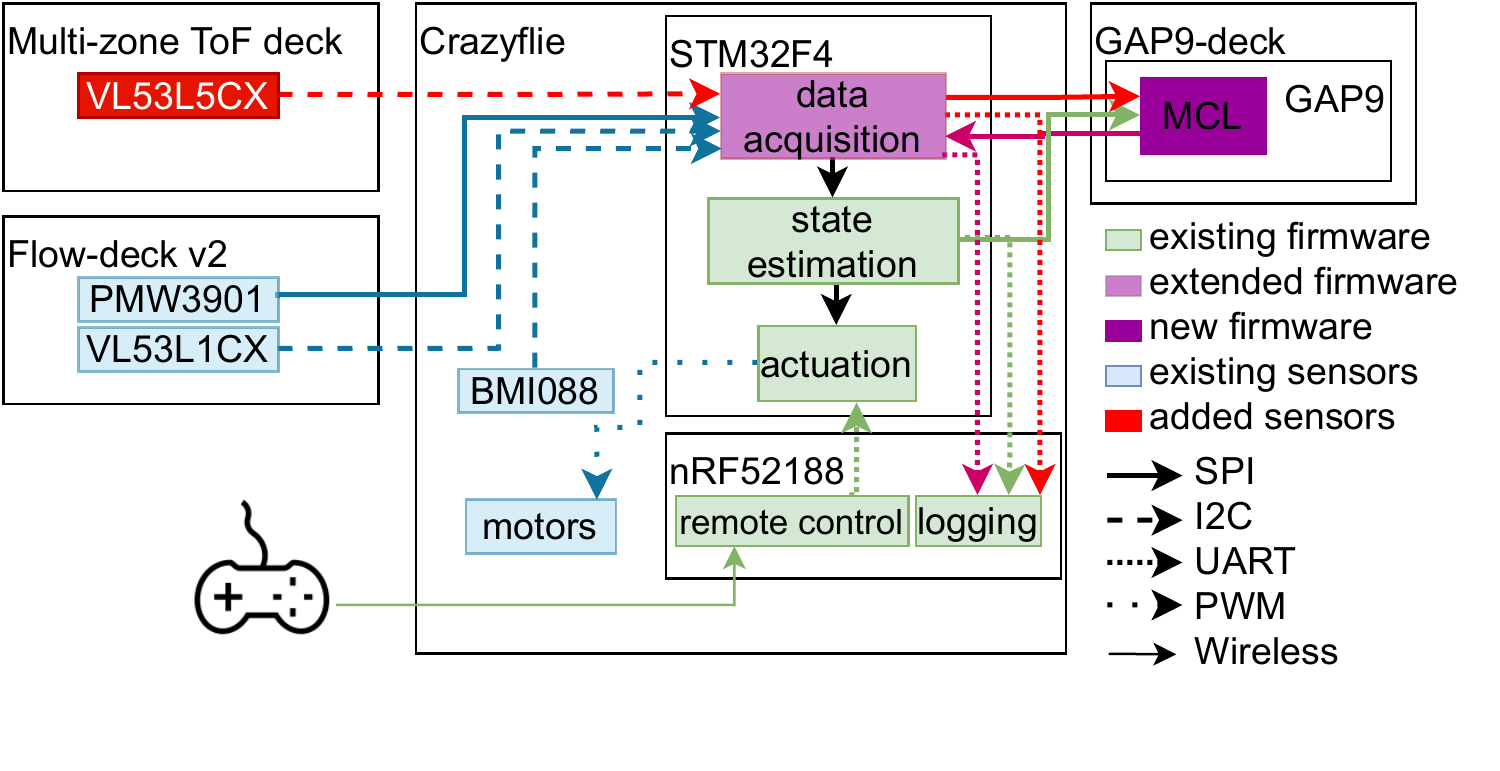}
  	\vspace{-1.0cm}
  \caption{System overview, showing the hardware connections and data dependencies between the Crazyflie and the three mounted extension decks.}
  \label{fig:sysoverview}
  	\vspace{-0.7cm}
\end{figure}

This section presents a complete description of the proposed infrastructure-less localization system for nano-UAVs; from the hardware design, to the algorithm implementation, and the in-field evaluation. We used the commercially available Crazyflie 2.1 platform from Bitcraze, extending its functionality with custom expansion boards featuring new sensors and processors, namely the VL53L5CX from STMicroelectronics and GAP9 SoC from GreenWaves technologies as main processing unit. All used components are commercially available, and our hardware design as well as the proposed embedded algorithm implementation will be released as open-source\footnote{\label{note:github}\url{https://github.com/ETH-PBL/Matrix_ToF_Drones}}. 
Fig.~\ref{fig:sysoverview} presents our design, composed of the Crazyflie's integrated hardware and software parts (blue for hardware, green for software) and our own additions (red for hardware, purple for software).
\vspace{-0.3cm}
\subsection{Hardware: Crazyflie and Extension Boards}
\vspace{-0.1cm}
The Crazyflie 2.1 is a commercially available open software/hardware nano-UAV. In this work, we use its inertial measurement unit (IMU), radio communication (using an nRF51822, solely to log data and steer the drone) and the main processor, an STM32F405 (\SI{168}{\mega\hertz}, \SI{192}{\kilo B}RAM), which is responsible for sensor readout, state estimation and real-time control. 
\subsubsection{Flow-deck v2} a commercially available deck, featuring a downward-facing optical flow sensor and 1D ToF sensor for odometry measurements. Those sensors improve the internal position estimate provided by the Crazyflie firmware through an extended Kalman Filter.
\subsubsection{Multizone-ToF-deck} a custom deck, featuring up to two VL53L5CX sensors (forward and backward facing), which can provide a matrix of either 8x8 or 4x4 pixels at maximally \SI{15}{\hertz} or \SI{60}{\hertz} respectively. For each zone, it provides a distance measurement coupled with an error flag, which gets raised when out of range measurements or interference are detected. 
\subsubsection{GAP9-deck} a custom deck, featuring GAP9, a RISC-V PULP-based SoC.
In our application, the multizone ToF sensor measurements are acquired by the STM32 via an I2C bus and then, together with the state estimation, sent on via SPI to the GAP9 SoC.
\vspace{-0.1cm}
\subsection{Processor: GAP9}
GAP9's architecture is based on the open-source SoC Vega~\cite{rossi2021vega} and features 10 RISC-V instruction set architecture-based cores, extended with custom instructions. The compute cluster, featuring 9 cores, one for orchestration and 8 workers, delivers programmable compute power at extreme energy efficiency. GAP9 features \SI{128}{\kilo B} of shared L1 memory. The fabric controller (FC) has access to various peripherals and features \SI{64}{\kilo B} RAM, \SI{1.5}{\mega B} interleaved memory (L2) and even \SI{2}{\mega B} flash.
The architecture employs adjustable dynamic frequency and voltage domains, allowing us to tune the energy consumption to the exact requirements at a particular point in time. At peak performance the cores run at \SI{400}{\mega\hertz} on both the cluster and the FC.
\vspace{-0.1cm}
\subsection{Algorithm: Monte Carlo Localization}\label{sec:mcl}
This section first provides an overview of the Monte Carlo localization algorithm and then explains our adaptions for running it on-board and in real-time on GAP9.
\subsubsection{Algorithm overview}
\begin{figure}[t]
  \centering
  \includegraphics[width=0.9\linewidth]{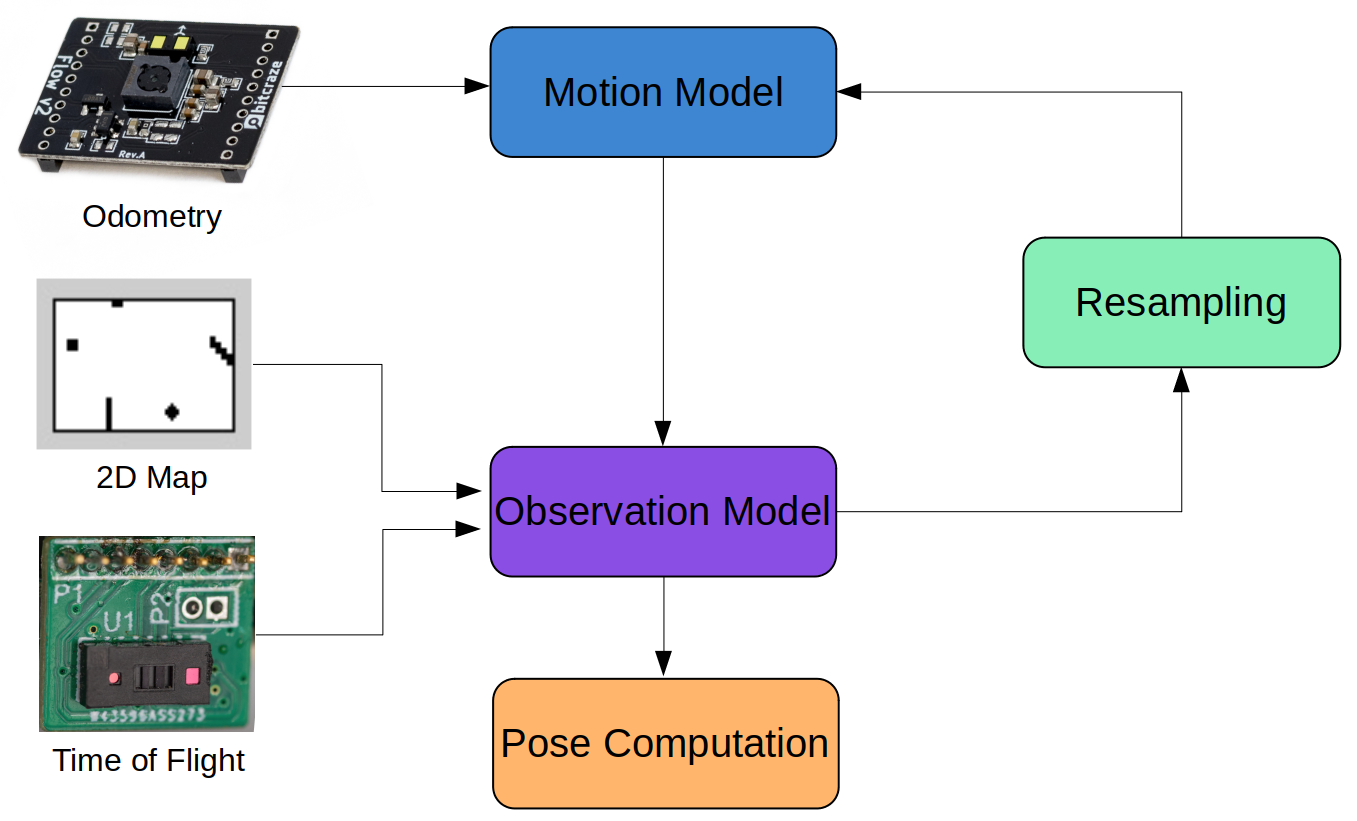}
    	\vspace{-0.2cm}
  \caption{The logic flow of the Monte Carlo localization algorithm.}
  \label{fig:mclflow}
  	\vspace{-0.5cm}
\end{figure}

MCL, most commonly used with occupancy grid maps~\cite{moravec1989sdsr,meyerdelius2021aaai,dellaert1999icra}, is a particle filter-based approach for estimating the posterior of the robot's pose $x_t$ given a map $m$, sensor readings $z_t$ and odometry inputs $u_t$. As the nano-UAV flies at a fixed height and localizes in a 2D grid map, the nano-UAV's state $x_t$ is defined by the 2D coordinates $(x,y)^\top$ and the yaw-angle orientation $\theta \in [0,2\pi)$. MCL has 3 main components: the prediction step using the motion model, the correction step using the observation model and resampling (\figref{fig:mclflow}). When odometry is available, we sample from the proposal distribution $p(x_t \mid x_{t-1}, u_t)$ with odometry noise $\sigma_{\text{odom}} \in \mathbb{R}^3$.  
The observation model describes the probability of observing $z_t$ from pose $x_t$ given a map $m$, where each observation ${z_t}$ is composed of $K$ elements $z_t^k$. As we are using a range sensor with an occupancy grid map, we chose the beam end point model~\cite{thrun2005probrobbook} 
as our observation model, as shown in Eq.~\ref{eq:beam_end_model},
  	\vspace{-0.2cm}
\begin{align}
 p(z_t^k\mid x_t,m) &=  \frac{1}{\sqrt{2 \pi \sigma_{\text{obs}}}} \exp{\left(-\frac{EDT(\hat{z}_t^k)^2}{2 \sigma_{\text{obs}}^2}\right)},
 \label{eq:beam_end_model}
 	\vspace{-0.5cm}
\end{align}
where $\hat{z}_t^k$ is the end point of the ToF beam in the occupancy grid map~$m$. 
We estimate the distance between each cell in the occupancy grid map to an obstacle (occupied cell) using the Euclidean distance transform~(EDT)~\cite{felzenszwalb2012toc}. The EDT is truncated at $r_{\text{max}}$, a predefined maximal range. In addition to the three main components~(\figref{fig:mclflow}), we also include a fourth step, pose computation, where the pose estimation is computed as the weighted average over all particles.

\subsubsection{Adaptions for Lightweight and Parallel Embedded Implementation}
The two main constraints in the nano-UAVs hardware are memory and time - we need to use both resources efficiently to enable MCL on-board in real-time. 
Our implementation of MCL is asynchronous -- the motion model is sampled when odometry is available, and the particles are re-weighted when new range measurements arrive. We only consider new observations if the drone moves more than $d_{xy}$ or rotates more than  $d_{\theta}$. However, we configured our sampling rates for the motion and observation update to be the same, limited by the \SI{15}{\hertz} maximum update rate of the ToF sensor. 

The motion model, observation model and pose computation can be parallelized exploiting the GAP9 cluster by distributing the particles among the cores. The resampling step can also be parallelized, however, as it depends on all weights, we can not plan the workload distribution optimally.

The first step is weight normalization, which involves  computing the sum and dividing by it -- we can parallelize this step by splitting the particles evenly to all cores. We also store the partial sums, as we can use them to parallelize the main resampling step, drawing the new particles.

\begin{figure}[t]
  \centering
  \includegraphics[width=0.65\linewidth]{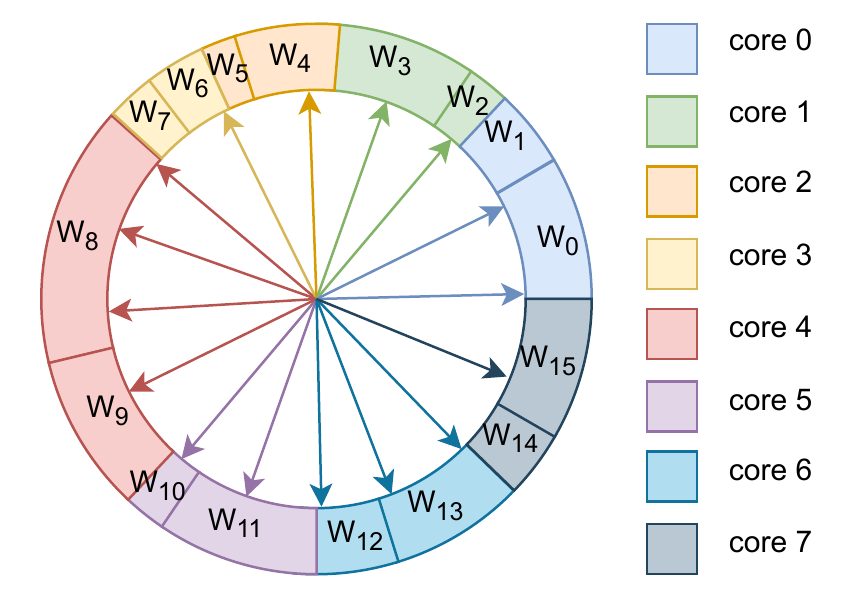}
  \caption{Parallelizing the resampling wheel: Each color represents a core, the current particles are distributed evenly (here two per core) and then the new particles are chosen according to the where the arrows of the resampling wheel point.}
  \label{fig:resamplingwheel}
  	\vspace{-0.5cm}
\end{figure}

For drawing the new particles, we use a systematic resampling algorithm~\cite{douc2005comparison}, which we explain with the analogy of a wheel, as shown in Fig. \ref{fig:resamplingwheel}. We draw one random number, corresponding to the first \textit{arrow} in the wheel, with the other $N-1$ \textit{arrows} being fixed at regular
intervals from that randomly picked \textit{arrow}.
The colors show how we distribute the drawing of the next particles to the cores. The current particle weights are represented by the cone area they occupy. As we know the partial sums computed by all cores, we can directly use them to calculate which core will resample how many particles and which. In  Fig. \ref{fig:resamplingwheel} the colored \textit{arrows} represent the new particles picked by the corresponding cores.

The main components of MCL using memory space are the particles and the map. The occupancy map requires 2 bits per cell (to represent the 3 possible states), to simplify the memory access we store it as 1 byte per cell. However, we also precompute the EDT values for each cell, leading to an additional floating point number being saved for every cell. To decrease the memory usage, we compare three possibilities: 32-bit floating point numbers, 16-bit floating point numbers and quantized 8-bit unsigned integer values. 
For the particles, we need four numbers each -- one for x position, y position, yaw angle and weight. With a 32-bit floating point representation, this leads to 16 bytes per particle. However, as we are double-buffering the particles for executing the resampling step, we need 32 bytes per particle for the 32-bit representation, 16 bytes for more memory-efficient 16-bit representations.

\section{Experimental Evaluation}
This section present our experiments to demonstrate the effective capabilities of our framework. The results of our experiments also support our key claims, our system can:
(i) accurately localize a nano-UAV in a given map, 
using low element-count sensors without infrastructure,  (ii) reduce memory consumption with quantization and lower precision floats without a significant drop in accuracy, (iii) reduce latency by a factor of 7 through parallelized implementation and localize on-board in real-time, and (iv) operate with low power consumption, where sensing and processing only consume below 7\% of the overall power.
 
\begin{figure}[t]
  \centering
  \includegraphics[width=0.95\linewidth]{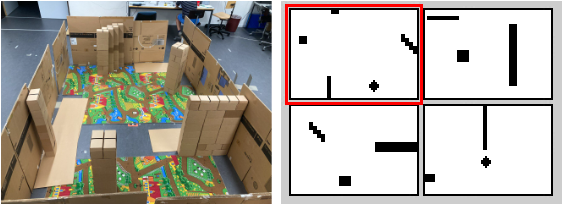}
  \caption{Left: The drone maze. Right: The occupancy grid map used for the localization task, where the highlighted part corresponds to the drone maze}
  \label{fig:mapmaze}
  	\vspace{-0.5cm}
\end{figure}

\vspace{-0.2cm}
\subsection{Experimental Setup}
To evaluate the performance of our approach, we recorded a dataset, including 6 sequences, while flying the drone in our "drone maze" (\figref{fig:mapmaze}). The recordings include ToF measurements from two sensors, internal state estimation based on the FlowDeck's optical flow and ground truth pose. The ground truth is extracted using a motion capture system, Vicon Vero 2.2, with six cameras positioned around the maze, covering an area of \SI{16}{\meter\squared}. 
The map acquisition is done by manually measuring the maze objects, which introduces some inaccuracy and increases the localization challenge. For all experiments, we use a map resolution of \SI{0.05}{\meter} by \SI{0.05}{\meter} per cell. The algorithm parameters are $\sigma_{\text{odom}}=$ (0.1\,m, 0.1\,m, 0.1\,rad), $\sigma_{\text{obs}}=2.0$, $r_{\text{max}}=$ 1.5\,m,  $d_{\text{xy}}=$ 0.1\,m and $d_{\theta}=$0.1\,rad,.
To challenge localization even further, we extended the map with 3 artificial mazes, to a total of \SI{31.2}{\meter\squared} of structured area. 

Three aspects were considered --- the localization accuracy, the runtime performance and the system power consumption. To evaluate the accuracy, we take into account 3 metrics - the success rate, the time to convergence and absolute trajectory error~(ATE) after convergence. We define convergence to occur when the estimated pose is close to the ground truth pose, within a distance of (\SI{36}{\degree}/\SI{0.2}{\meter}). The localization is counted as successful if the pose tracking remains reliable from convergence until the end of the sequence, meaning that the ATE does not exceed \SI{1}{\meter}.

\subsection{Robust Localization with Sparse ToF}
To support the claim that we are able to accurately localize with low element-count sensors, we evaluated our approach on six sequences. For each sequence, we repeated the localization experiments with six different random seeds to verify robustness. All of our experiments were executed using data from both, front and rear, ToF sensors, unless explicitly mentioned. 

\begin{figure}[t]
  \centering
  \includegraphics[width=1.0\linewidth]{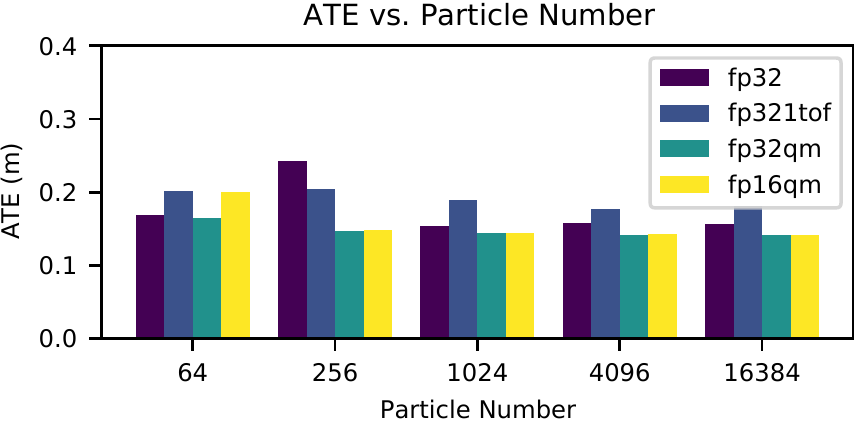}
  	\vspace{-0.7cm}
  \caption{The absolute trajectory error (ATE) computed over all sequences.}
  \label{fig:atevsmethod}
  	\vspace{-0.5cm}
\end{figure}

\begin{figure}[t]
  \centering
  \includegraphics[width=1.0\linewidth]{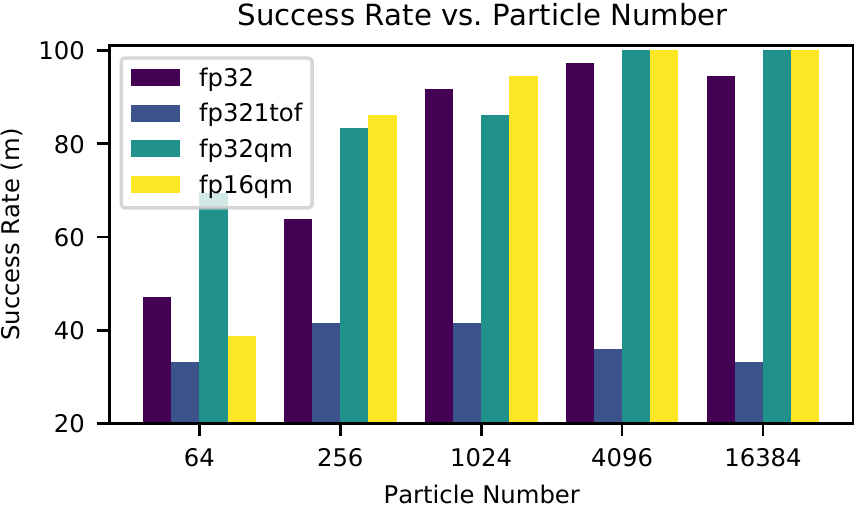}
  	\vspace{-0.7cm}
  \caption{The success rate in $\%$ of successful convergence across all sequences.}
  \label{fig:successvsmethod}
  	\vspace{-0.6cm}
\end{figure}

Infrastructure-dependent localization approaches~\cite{niculescu2022iotj,van2020board} for nano-UAVs, which were evaluated in similar environments to ours, have achieved mean localization errors of \SI{0.22}{\meter} and \SI{0.28}{\meter}. As can be seen in \figref{fig:atevsmethod} and \figref{fig:successvsmethod}, our approach can localize with \SI{0.15}{\meter} accuracy and achieves above $95\%$ success rate with sufficient number of particles, outperforming the existing approaches. An illustration of successful localization can be seen in \figref{fig:motivation}. Our experiments show that our approach is robust with respect to the number of particles, providing ATE of less than \SI{0.2}{\meter} for a large range of particle numbers.

\begin{figure}[t]
  \centering
  \includegraphics[width=1.0\linewidth]{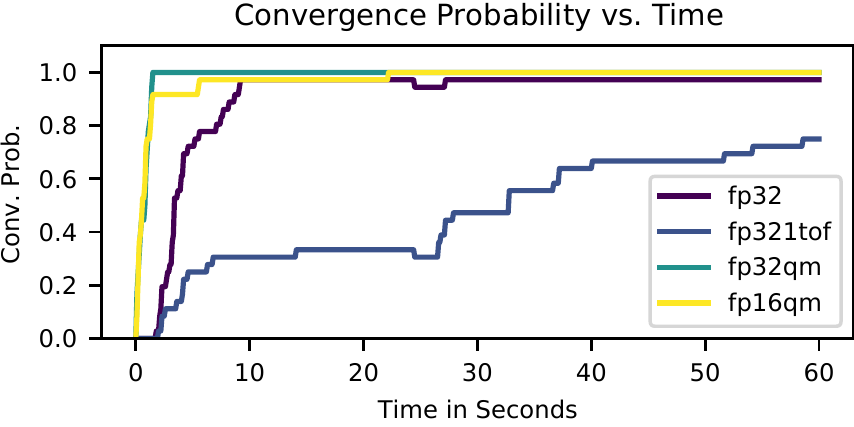}
  	\vspace{-0.7cm}
  \caption{The probability of converging over time, across all sequences, computed for 4096 particles.}
  \label{fig:convprob}
  	\vspace{-0.5cm}
\end{figure}

We perform additional experiments to confirm the contribution of the second (rear) ToF sensor. We compare the performance with a pair of ToF sensors, refer to as fp$32$, to that of a single ToF sensor which we refer to as fp$321$tof. For both configurations, the accuracy is calculated for a 32bit representation of the floats in the EDT and a particle's weight and pose.
As shown in \figref{fig:successvsmethod}, the success rate when using two ToF sensors is significantly higher, and the accuracy is also improved sightly~(\figref{fig:atevsmethod}). The convergence is slower when using only 1 ToF sensor, as illustrated in \figref{fig:convprob}.

\vspace{-0.2cm}
\subsection{Memory Usage}
To support our claim, that we can reduce memory consumption with quantization and lower precision floats without significant loss of accuracy, we compare our optimized implementation against a full-precision implementation. 
We compare the localization performance for three implementations of our approach. 
The first implementation uses a 32-bit representation for the floats in the EDT and particles (fp$32$). The second implementation, (fp$32$qm) is using a quantized EDT whose values are 8-bit unsigned integers. The third implementation (fp$16$qm)  is a quantized EDT and an 16-bit float representation for a particle's weight and pose. 
As shown in \figref{fig:atevsmethod} and \figref{fig:successvsmethod}, the quantized implementations maintain high success rate and provide accuracy that surpasses the full precision implementation. As reported in \figref{fig:convprob}, the convergence time is improved for the optimized implementations. We speculate that the quantization accelerates the rate in which weak particles are eliminated in the resampling step, leading to faster convergence and overall better performance. 

\begin{figure}[t]
  \centering
  \includegraphics[width=0.95\linewidth]{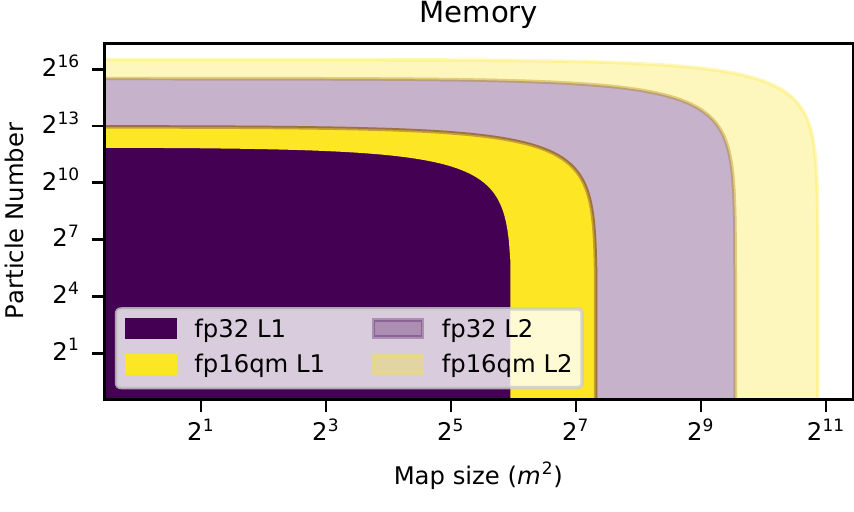}
  	\vspace{-0.4cm}
  \caption{The trade off between the number of particles and the map size at a resolution of \SI{0.05}{\meter}/cell for both L1 and L2 memory.}
  \label{fig:memory}
  	\vspace{-0.4cm}
\end{figure}

As we do not observe a significant accuracy loss when going down to 8-bit quantized EDT values, we can reduce our memory requirements for the map from 5 bytes per cell to 2 bytes per cell. 
For the particles, we saw that using half precision instead of full precision representation does not result in a significant accuracy drop, meaning we can reduce the needed memory by a factor of 2.
We visualize the memory savings in Fig.~\ref{fig:memory}, where we show how many particles and square meters can be stored on the GAP9 in L1 (blue full precision, yellow quantized/FP16) respectively L2 (red full precision, green quantized/FP16) memory.
\vspace{-0.2cm}
\subsection{Real-time Performance}

In support of our third claim, that we can reduce latency through parallelized implementation and accurately localize
on-board in real-time, we compared our optimized implementation against a naive implementation, and measured the latency improvement achieved by parallelization. 

\begin{figure}[t]
  \centering
  \includegraphics[width=0.9\linewidth]{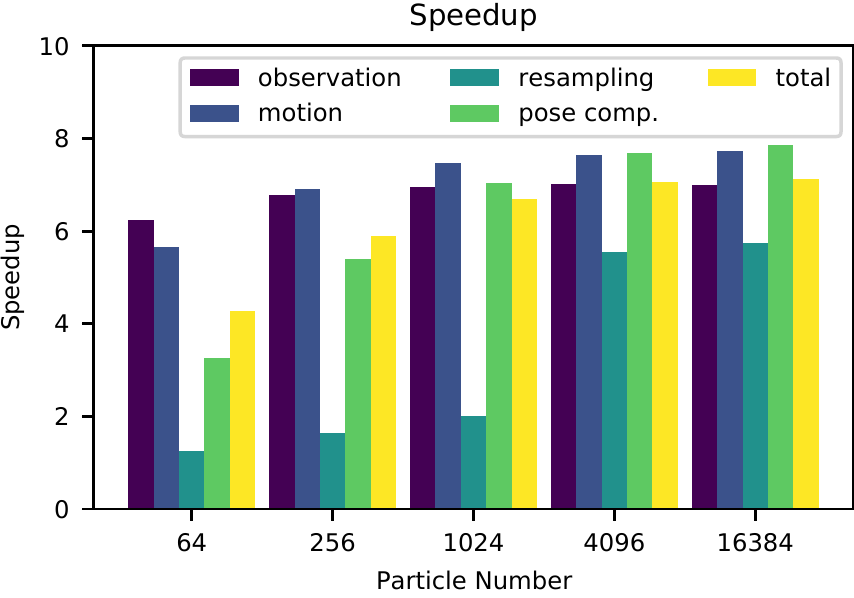}
  	\vspace{-0.2cm}
  \caption{The achieved speedup on GAP9 by parallelizing the correction step, motion update, resampling step and prediction computation for different numbers of particles.}
  \label{fig:speedup}
  	\vspace{-0.5cm}
\end{figure}

In Tab.~\ref{tab:exectimes} we report the execution times per particle for the four previously described steps executed sequentially on one core and executed in parallel on 8 cores (while using the ninth core of the cluster for orchestration). In Fig.~\ref{fig:speedup} we show the achieved speedup for different numbers of particles. As expected, the resample step scales the worst - however, for high numbers of particles we can reach more than 5x speedup even for this step. We also report the total speedup achieved in orange - we can observe it improving until a factor of 7 at high numbers of particles. Note that the total execution time is around \SI{40}{\micro\second} higher than the sum of the four tasks, independent of the numbers of particles and multicore usage, which are used for preprocessing the sensor data and transferring information to the tasks.

\begin{table}[t]
  \caption{Execution times by components for different numbers of particles}
   	\vspace{-0.2cm}
   \centering
   \resizebox{\columnwidth}{!}{
 \begin{tabular}{cccccc}\toprule
 &\multicolumn{5}{c}{execution time per particle 1 core / 8 cores in \SI{}{\nano\second}, GAP9@\SI{400}{\mega\hertz}} \\ \midrule

Particles & 64 & 256 & 1,024 & 4,096$^{\mathrm{a}}$ & 16,384$^{\mathrm{a}}$ \\  \midrule

Observation & 8531/1412 & 8484/1313 & 8518/1283 & 8649/1294 & 8704/1295 \\
Motion & 2828/500 & 2715/391 & 2689/357 & 3002/390 & 2985/386 \\
Resampling & 313/250 & 191/121 & 161/84 & 558/108  & 556/104 \\
Pose Comp. & 750/234 & 633/117 & 604/86 & 777/101 & 775/99 \\
  \bottomrule
 \end{tabular}
 }
 \label{tab:exectimes}
 \\
\flushleft
\footnotesize{$^{\mathrm{a}}$Particles stored in L2.}
	\vspace{-0.6cm}
\end{table}

\begin{table}[t]
  \caption{Average power consumption of the MCL algorithm on GAP9 at different operating points.}
     	\vspace{-0.2cm}
   \centering
  \resizebox{\columnwidth}{!}{\begin{tabular}{lcc}\toprule
& Avg. power consumption & Execution time \\ \midrule
GAP9@\SI{400}{\mega\hertz}/1,024 particles & \SI{61}{\milli\watt}& \SI{1.901}{\milli\second}\\
GAP9@\SI{12}{\mega\hertz}/1,024 particles & \SI{13}{\milli\watt} & \SI{59.898}{\milli\second}\\
GAP9@\SI{400}{\mega\hertz}/16,384 particles$^{\mathrm{a}}$ & \SI{61}{\milli\watt} & \SI{30.880}{\milli\second}\\
GAP9@\SI{200}{\mega\hertz}/16,384particles$^{\mathrm{a}}$ & \SI{38}{\milli\watt} & \SI{61.524}{\milli\second}\\
  \bottomrule
 \end{tabular}
 }
 \label{tab:mclpower}
\\
\flushleft
\footnotesize{$^{\mathrm{a}}$Particles stored in L2.}
	\vspace{-0.6cm}
\end{table}

\vspace{-0.2cm}
\subsection{Power Consumption}
In support of our fourth claim, that our approach can operate with low power consumption, we performed power measurements. 
Firstly, we analyzed the power consumption for processing on GAP9. We analyze two working points, 1,024 particles, which can still fit in L1, and 16,384 particles, the maximum number of particles we considered. We measured the power consumption at the maximum possible frequency and at the minimal frequency at which we can still operate in real-time, meaning processing in less than \SI{67}{\milli\second}. In \tabref{tab:mclpower}, we report our results, consisting of the average power consumption and the execution times at the respective frequencies. 

Combined with the power needed for the sensors, \SI{320}{\milli\watt} each, and the remaining Crazyflie electronics (besides motors), \SI{280}{\milli\watt}, all sensing and processing power sums to \SI{981}{\milli\watt}. This is around 7\% of the overall power consumption of the drone.
\vspace{-0.3cm}
\section{Conclusion}
This paper presents a system which includes a nano-UAV, a low-power multi-core processor, low element-count sensors, and a hardware-specific highly-optimized MCL implementation for indoor localization. Our experiments show that this system enables nano-UAVs to accurately localize in a given map despite the sparsity of the sensors. The paper also demonstrates the benefit of the proposed parallelized and memory-efficient MCL, which runs at \SI{15}{\hertz}, efficiently using the available compute resources to operate onboard, in real-time, without significant loss of accuracy. Sensing and processing, even in the most powerful configuration, increases the drone's power consumption only by 7\%. Future works will extend the proposed system to applications such as path planning and exploration. 

\vspace{-0.2cm}
\section*{Acknowledgment}
The authors thank V.Niculescu for the picture of the drone.
\vspace{-0.3cm}
\bibliographystyle{IEEEtran}
\vspace{-0.3cm}
\bibliography{IEEEabrv,new}

\end{document}